\documentclass[trackchanges,twocolumn,twocolappendix]{aastex701}

\usepackage[version=4]{mhchem}
\usepackage{graphicx}
\usepackage{subcaption}
\usepackage{soul}
\usepackage{tabularx}
\usepackage{multirow}%
\usepackage{booktabs} 
\usepackage{seqsplit}
\usepackage{threeparttable}
\usepackage{bm}
\usepackage{placeins}   
\usepackage{float}      
\usepackage{upgreek}
\allowdisplaybreaks     

\begin{document}

\title{C, N, O, S, and photochemistry in a temperate giant planet orbiting a late M dwarf}

\author[orcid=0000-0002-0659-1783]{Michael Zhang}
\altaffiliation{51 Pegasi b Fellow}
\affiliation{Department of Astronomy and Astrophysics, University of Chicago, Chicago, IL 60637, USA}
\email[show]{mzzhang2014@gmail.com}

\author[orcid=0000-0002-6215-5425]{Qiao Xue}
\affiliation{Department of Astronomy and Astrophysics, University of Chicago, Chicago, IL 60637, USA}
\email{qiaox@uchicago.edu}

\author[orcid=0000-0002-1551-2610]{Jeehyun Yang}
\affiliation{Department of Astronomy and Astrophysics, University of Chicago, Chicago, IL 60637, USA}
\email{jeehyuny@uchicago.edu}

\author[orcid=0000-0001-5909-4433]{Vighnesh Nagpal}
\altaffiliation{NSF Graduate Research Fellow}
\affiliation{Department of Astronomy and Astrophysics, University of Chicago, Chicago, IL 60637, USA}
\email{vnagpal@uchicago.edu}

\author[orcid=0000-0002-2338-476X]{Michael R. Line}
\affiliation{School of Earth and Space Exploration, Arizona State University, Tempe, AZ 85287, USA}
\email{Michael.Line@asu.edu}

\author[orcid=0000-0002-3263-2251]{Guangwei Fu}
\affiliation{Department of Physics and Astronomy, Johns Hopkins University, Baltimore, MD, USA}
\email{guangweifu@gmail.com}

\author[orcid=0000-0001-8236-5553]{Matthew C.\ Nixon}
\altaffiliation{51 Pegasi b Fellow}
\affiliation{School of Earth and Space Exploration, Arizona State University, Tempe, AZ 85287, USA}
\email{mcnixon1@asu.edu}

\author[orcid=0000-0003-4733-6532]{Jacob L.\ Bean}
\affiliation{Department of Astronomy and Astrophysics, University of Chicago, Chicago, IL 60637, USA}
\email{jacobbean@uchicago.edu}

\author[orcid=0000-0002-8518-9601]{Peter Gao}
\affiliation{Earth and Planets Laboratory, Carnegie Institution for Science, 5241 Broad Branch Road, NW, Washington, DC 20015, USA}
\email{pgao@carnegiescience.edu}

\author[orcid=0000-0002-1337-9051]{Eliza M.-R. Kempton}
\affiliation{Department of Astronomy and Astrophysics, University of Chicago, Chicago, IL 60637, USA}
\email{ekempton@uchicago.edu}

\author[orcid=0000-0003-0156-4564]{Luis Welbanks}
\affiliation{School of Earth and Space Exploration, Arizona State University, Tempe, AZ 85287, USA}
\email{luis.welbanks@asu.edu}

\author[orcid=0000-0001-7904-4441]{Edward M. Bryant}
\affiliation{Department of Physics, University of Warwick, Gibbet Hill Road, Coventry CV4 7AL, UK}
\affiliation{Centre for Exoplanets and Habitability, University of Warwick, Gibbet Hill Road, Coventry CV4 7AL, UK}
\email[]{edward.m.bryant@warwick.ac.uk}


\author[orcid=0000-0001-6023-1335]{Daniel Bayliss}
\affiliation{Department of Physics, University of Warwick, Gibbet Hill Road, Coventry CV4 7AL, UK}
\email[]{d.bayliss@warwick.ac.uk}

\author[orcid=0000-0003-2404-2427]{Madison Brady}
\affiliation{Department of Physics \& Astronomy, Michigan State University, East Lansing, Michigan, 48824}
\email{bradym27@msu.edu}

\author[orcid=0000-0002-0875-8401]{Jean-Michel D\'esert}
\affiliation{Leibniz Institute for Astrophysics, AIP Potsdam, Potsdam, 14482 Potsdam, Germany}
\affiliation{DESY, Platanenallee 6, Zeuthen, D-15738, Germany}
\email[]{jmdesert@aip.de}

\author[orcid=0000-0001-5542-8870]{Vincent Van Eylen}
\affiliation{Mullard Space Science Laboratory, University College London, Holmbury St Mary, Dorking, Surrey, RH5 6NT, UK}
\email[]{v.vaneylen@ucl.ac.uk}

\author[orcid=0000-0002-9843-4354]{Jonathan J. Fortney}
\affiliation{Department of Astronomy and Astrophysics, University of California, Santa Cruz, CA 95064, USA}
\email[]{jfortney@ucsc.edu}

\author[orcid=0000-0002-5389-3944]{Andr\'es Jord\'an}
\affiliation{Facultad de Ingenier\'ia y Ciencias, Universidad Adolfo Ib\'{a}\~{n}ez, Av. Diagonal las Torres 2640, 7941169 Pe\~{n}alol\'{e}n, Santiago, Chile}
\affiliation{Departamento de Astronomía, Universidad de Chile, Casilla 36-D, Santiago, Chile}
\email{ajordan@astrofisica.cl}

\author[orcid=0000-0001-9521-6258]{Vivien Parmentier}
\affiliation{Laboratoire Lagrange, Université de la Côte d’Azur, Observatoire de la Côte d’Azur, CNRS, Nice, France.}
\email[]{vivien.parmentier@oca.eu}

\author[orcid=0000-0002-2875-917X]{Caroline Piaulet-Ghorayeb}
\altaffiliation{E. Margaret Burbidge Postdoctoral Fellow}
\affiliation{Department of Astronomy and Astrophysics, University of Chicago, Chicago, IL 60637, USA}
\email[]{carolinepiaulet@uchicago.edu}

\author[orcid=0000-0002-7444-5315]{Elyar Sedaghati}
\affiliation{European Southern Observatory (ESO), Alonso de C\'{o}rdova 3107, 763 0355 Santiago, Chile}
\email{esedagha@eso.org}

\author[orcid=0000-0002-7352-7941]{Kevin B. Stevenson}
\email{Kevin.Stevenson@jhuapl.edu}
\affiliation{Johns Hopkins University Applied Physics Laboratory, 11100 Johns Hopkins Rd., Laurel, MD 20723, USA}

\author[orcid=0000-0002-5510-8751]{Amaury H.M.J. Triaud}
\affiliation{School of Physics \& Astronomy, University of Birmingham, Edgbaston, Birmingham B15 2TT, United Kingdom}
\email[]{a.triaud@bham.ac.uk}

\begin{abstract}
We report the JWST NIRSpec/PRISM transit spectrum of TOI-6894~b, an exceptional 420\,K sub-Saturn that is one of the rare giant planets transiting a late M dwarf.  Remarkably, both the light curve and the transit spectrum exhibit almost no stellar contamination.  The spectrum is dominated by prominent absorption features from CH$_4$ and the photochemical product CS$_2$.  For the first time in a transit spectrum, NH$_3$ is visually evident, while subtler features from H$_2$O and CO$_2$ can also be seen.  We significantly improve upon state-of-the-art photochemical reaction networks, and use our new network to run radiative-convective photochemical (``RCP'') models at different metallicities.  These models show that the spectrum--in particular the size of the NH$_3$ and CO$_2$ features relative to the CH$_4$ and H$_2$O features--is most consistent with a metallicity of 3--10$\times$ solar.  Using a semi-free retrieval framework that perturbs the RCP model's abundance and temperature profiles to fit the data, we find that the planet's C/O, N/O, and S/N ratios are consistent with solar values.  A grid retrieval on 1D radiative-convective photochemical equilibrium (RCPE) models reveals a similar result: $[M/H]=0.46 \pm 0.08$ and C/O=$0.69 \pm 0.06$.  The planet's atmospheric metallicity, abundance ratios, and bulk metal fraction are all strikingly similar to that of Jupiter, Saturn, and other gas giant exoplanets, despite orbiting a very low-mass star.
\end{abstract}



\section{Introduction} \label{sec:intro}
Giant exoplanets orbiting M stars (``GEMS'') are rare, with an occurrence rate of only $0.14 \pm 0.10$\% for stars between 0.088--0.26 $M_\odot$ \citep{bryant_2023}.  These planets' large planet-to-stellar mass ratios challenge the core accretion theory of planet formation.  At the same time, their large transit depths provide an excellent opportunity to examine the atmospheres --- and potentially unravel the mysteries --- of these rare enigmatic objects.  Despite their favorable observability, the first published JWST observations of GEMS targets have been difficult to interpret due to heavy stellar contamination \citep{ashtari_2026, kanodia_2026, canas_2026}.

The most extreme GEMS is TOI-6894~b, a puffy sub-Saturn ($R_p=0.86 R_J$, $M_p=0.17 M_J$) on a 3.4 day orbit around a M5 dwarf ($R_*=0.23 R_\odot, M_*=0.21 M_\odot, T_{\rm eff}=3000 \mathrm{K}$) that is the lowest mass star known to host a transiting giant planet \citep{bryant_2025}.  The $\sim$420 K TOI-6894~b offers a rare opportunity: the low gravity of the planet and the small size of the host star give it exceptionally large spectral features of 1300 ppm per scale height and a transmission spectroscopy metric \citep[TSM;][]{kempton_2018} of 450.  

Probing the chemical inventory of exoplanet atmospheres in different regions of parameter space has long been a goal of exoplanetary science.  Elemental abundances, in particular, reveal clues about planet formation (c.f.\ \citealt{oberg_2011}), although the interpretation is far from trivial \citep{feinstein_2025}.  TOI-6894~b is particularly well suited for this science because of its 420\,K equilibrium temperature --- substantially colder than all other giants orbiting M dwarfs that JWST has observed.  At this temperature, methane dominates the spectral features, gaseous water is present in observable quantities, nitrogen is observable as ammonia, and (as we will show) photochemistry converts the sulfur inventory originally locked in \ce{H2S} into \ce{CS2} \citep{moses_2024,mukherjee_2025,veillet_2026}, which has strong features at 4.3 and 4.7 $\mu$m.  Remarkably, then, TOI-6894~b offers the rare opportunity to probe the elemental inventories of C, N, O, and S in a single observation.

In this paper, we report the transit spectrum of TOI-6894~b as observed with JWST's NIRSpec instrument in PRISM mode.  We reduce the data with two independent pipelines (Section \ref{sec:obs}) and compare the spectrum with radiative-convective photochemical 
(RCP) models, assuming different metallicities (Section \ref{sec:modeling}).  To infer the elemental inventory, we use a combination of grid retrievals and ``semi-free'' retrievals based on perturbations of a single RCP model.  We discuss our inferred atmospheric properties as well as the bulk metallicity obtained from interior modelling in Section \ref{sec:discussion}, before concluding in Section \ref{sec:conclusion}.  In a companion paper \citep{yang2026cs2}, we discuss in detail the state-of-the-art photochemical model we developed for this letter, which is widely applicable to H$_2$/He-dominated planets at similar temperatures.

\section{Observation and Data Reduction}
\label{sec:obs}
We observed a single transit of TOI-6894~b with the Near-Infrared Spectrograph \citep[NIRSpec;][]{jakobsen_near-infrared_2022} on JWST on Jan 10, 2026 as part of program 8696 (PI: Michael Zhang). The instrument was configured in Bright Object Time Series (BOTS) mode with the PRISM disperser and the SUB512 subarray read out using the NRSRAPID readout pattern, providing low-resolution ($R \sim 100$) spectra spanning $0.6$--$5.3\,\mu\mathrm{m}$. The observation comprised 7177 integrations of 10 groups each, for a total duration of 5.0\,h, fully covering the 1.4\,h transit together with a sufficient out-of-transit baseline before and after the event.  The brightest non-anomalous pixel reaches $\sim35,000$ DN in the final group, well below the saturation limit of 61,000--64,000 DN.  All the JWST data used in this paper can be found in MAST: \dataset[10.17909/aqn9-zt52]{http://dx.doi.org/10.17909/aqn9-zt52}.

We reduced and analyzed the data with the \texttt{SPARTA} \citep{kempton_reflective_2023, xue_jwst_2025} and \texttt{Tswift} \citep{kirk_jwstnircam_2024} pipelines, which are completely independent of each other.  We provide full details for both reductions in Appendix~\ref{appendix:data}, but highlight some differences from each other and from standard methodology here. \texttt{SPARTA} uses optimal extraction, while \texttt{Tswift} uses box extraction.  Both use a linear-with-time systematics model, which we found to be statistically favored over both a quadratic and an exponential plus linear model.  We do not decorrelate against position because the transit is so deep that it substantially changes the observed stellar spectrum, giving rise to illusory position shifts in the wavelength direction (see Appendix Figure~\ref{appendix:figure_lightcurve}).
 
The main departure from standard methodology is in our limb darkening treatment.  Although the quadratic parameterization is most commonly used, the power-2 law matches stellar models more closely than the quadratic or any other two-parameter law used in the literature, especially for cool stars \citep{morello_2017}.  For our white light curve specifically, the power-2 parameterization results in a better fit ($\ln(B)=13$).  We therefore adopt this limb darkening law for our fiducial \texttt{SPARTA} reduction.  The \texttt{Tswift} reduction maintains the traditional quadratic parameterization, except the first coefficient is fixed to the value predicted by PHOENIX stellar models.

\begin{figure}[ht!] 
    \centering 
    \includegraphics[width=0.45\textwidth]{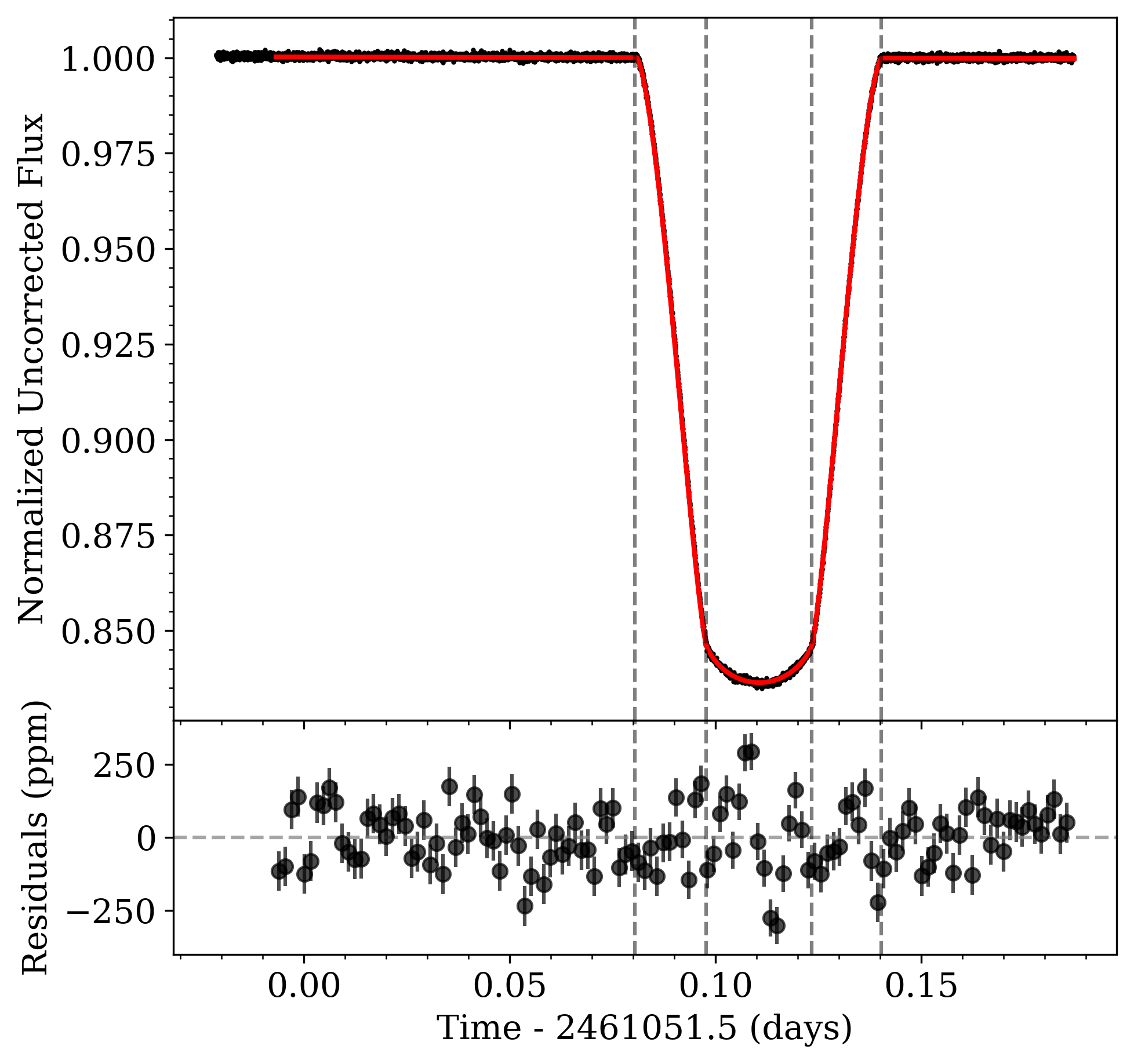} 
    \caption{The normalized, uncorrected white-light curve of TOI-6894~b spanning 0.8--5.2~$\mu$m, with the best-fit transit model overplotted in red. The first 500 integrations are discarded to get rid of detector settling effects. The vertical dashed lines indicate the times of ingress and egress. Binned residuals are shown below with the y-axis magnified by a factor of 76 relative to the upper panel.}
    \label{fig:lightcurve} 
\end{figure}

\begin{figure*}[ht!] 
    \centering 
    \includegraphics[width=1.0\textwidth]{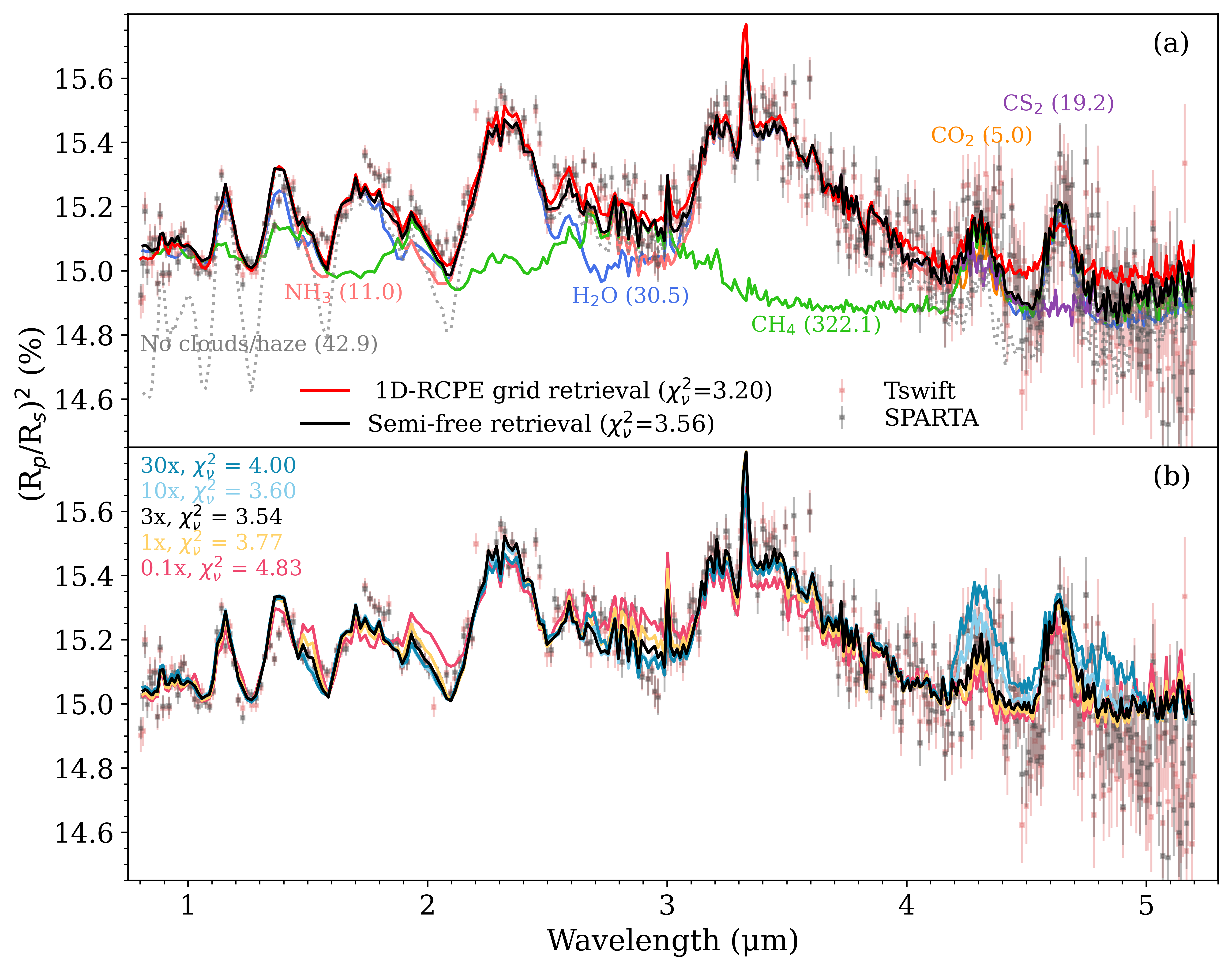} 
    \caption{Transmission spectra of TOI-6894~b from two independent reductions, compared to various models and their reduced $\chi^2$ values. We only considered 0.8--5.2 $\mu$m for modeling due to the sharp drop in PRISM throughput outside this range. (a) The \texttt{SPARTA} and \texttt{Tswift} spectra overlaid with the best-fit 1D-RCPE grid retrieval (solid red) and semi-free retrieval model (solid black). Coloured curves show the best-fit semi-free model with each individual molecule removed. For each species, we re-ran the retrieval with that molecule's opacity contribution muted and computed the log Bayes factor $\ln(B)= \ln Z_{full} - \ln Z_{mol,muted}$ using \texttt{PyMultiNest}'s global evidence. The resulting  $\ln(B)$ are indicated in parentheses next to each label. $\ln(B)>5$ usually indicates a strong preference for the model \citep{trotta_bayes_2008}. (b) RCP models assuming different metallicities (\S\ref{subsec:self_consistent}) are shown in colors, with their $\chi_{\nu}^2$ values listed in the legend. A clear preference for 3--10$\times$ solar metallicity over sub-solar or 30$\times$ solar is apparent by eye, where the 30$\times$ model overpredicts CO$_2$ (4.2--4.4 $\mu$m) along with OCS (4.8--5.0 $\mu$m), and the sub-solar to solar models overpredict NH$_3$ (1.95 $\mu$m, 1.5 $\mu$m, 2.95 $\mu$m) and underpredict CO$_2$. }
    \label{fig:spectra} 
\end{figure*}

The raw \texttt{SPARTA} white light curve is shown in Figure~\ref{fig:lightcurve}, and transmission spectra from both reductions are shown in Figure~\ref{fig:spectra}.  Both the light curve and the spectrum are remarkably clean.  In stark contrast to other GEMS that JWST has observed, there are no obvious flares or prominent starspot crossings in the light curve and no obvious transit light-source effect in the spectrum.  There are still sources of unexplained noise: the scatter in the light curve is 2.2$\times$ photon noise, and there is some visual evidence for correlated noise at the $\pm$100~ppm level out of transit.  Within transit, the correlated noise appears marginally worse (with $\pm 250$ ppm outliers), possibly due to stellar inhomogeneities.  

The two reductions yield transmission spectra that agree within $1\sigma$ in 95\% of the spectroscopic channels at the native PRISM resolution (Figure \ref{fig:spectra}).  The transit spectrum is dominated by CH$_4$, for which five absorption bands are visible by eye.  The prominent peak at 4.65 $\mu$m is due to the photochemical product CS$_2$.  CS$_2$ has a second, weaker absorption band at 4.3 $\mu$m, almost exactly overlapping the CO$_2$ band at that wavelength.  NH$_3$, long elusive and only faintly glimpsed in previous JWST transit spectra (e.g. \citealt{welbanks_2024}), now shows up clearly: it creates the small peak at 1.95 $\mu$m, the knee and plateau at 1.5 $\mu$m, and the long plateau around 2.95 $\mu$m.  H$_2$O has subtle features, serving mostly to fill in the plateau at 2.5--3.0 $\mu$m.  The extraordinary precision and richness of this spectrum, with absorption features from carbon, nitrogen, oxygen, and sulfur species, opens the door to atmospheric inferences rarely possible for other exoplanets.  In the following sections, we will merely scratch the surface of these possibilities.

\section{Modelling} \label{sec:modeling}
As a temperate planet, TOI-6894~b is far from chemical equilibrium.  Vertical mixing from the hot, high pressure interior is expected to decrease ammonia while favoring N$_2$, and decrease CH$_4$ while favoring CO \citep{moses_2013,fortney_2020}. The prominent \ce{CS2} feature highlights the importance of another disequilibrium process: photochemistry.  As a result, equilibrium models would not be appropriate for this planet.

To understand the temperature, cloud properties, and chemical inventory of TOI-6894~b, we use a range of models.  On the most physically motivated end are radiative-convective (\texttt{PICASO}) and photochemical (\texttt{EPACRIS}) models (Subsection \ref{subsec:self_consistent}), for which we developed a custom reaction network that is more systematically and comprehensively constructed than any other publicly available network.  Using a technique we call a semi-free retrieval (Subsection \ref{subsec:semi-free}), we perturb the temperature profile, vertical mixing ratio profiles, and other parameters from these RCP models to better match the data.  In this way, we can constrain atmospheric properties--including the elemental abundances of C, N, O, and S--while maintaining physical realism.  The final atmospheric modelling approach we use (Subsection \ref{subsec:schimera}) is a grid retrieval based on the 1D self-consistent radiative-convective model \texttt{ScCHIMERA} and the photochemical code \texttt{Photochem}.  All three methods lead to consistent atmospheric compositions, and therefore to consistent inferences about the bulk metallicity (Subsection \ref{subsec:interior_modelling}).

\subsection{Radiative-convective photochemical (RCP) forward models with PICASO and EPACRIS}
\label{subsec:self_consistent}
We perform 1D radiative-convective photochemical (RCP) modelling using a combination of \texttt{PICASO}, \texttt{EPACRIS}, and \texttt{PLATON}. \texttt{PICASO} computes the radiative-convective equilibrium (RCE), \texttt{EPACRIS} solves for the steady-state photochemical kinetic-transport composition based on the resulting temperature-pressure profile using \texttt{PICASO}, and \texttt{PLATON} performs the radiative transfer.  Because we do not iterate PICASO and EPACRIS to convergence, the model as a whole is not strictly self-consistent.

We carried out RCE calculations with \texttt{PICASO 4.0} \citep{mang2026picaso} assuming complete day--night heat redistribution\footnote{This corresponds to \texttt{rfacv = 0.5} \citep{mukherjee_picaso2023}}. The calculations were performed over 91 pressure layers spanning $10^{-6}$ to $10^{3}$~bar. We adopted $T_{\mathrm{int}} = 100$~K for our fiducial model, reflecting the likely advanced age of the quiet star. The incident flux at the top of the atmosphere was calculated by interpolating the \texttt{PHOENIX} grid of stellar models according to TOI-6894's stellar parameters, as described in \citep{mukherjee_picaso2023}.

Based on the $T$--$P$ profiles, we performed one-dimensional photochemical kinetic-transport modeling of various atmospheric scenarios of TOI-6894~b using \texttt{EPACRIS} \citep{yang_2024}. We adopted the photochemical network from \citet{yang2024chemical}. To account for \ce{CS2} chemistry, we additionally included \ce{CS}, \ce{CS2}, \ce{H2CS}, \ce{HCS}, and \ce{NS}, together with their associated chemical reactions (3 photochemical reactions--incorporating UV cross sections--and 36 thermochemical reactions) from VULCAN's \texttt{SNCHO\_photo\_network\_2025.txt} network \citep{VULCAN_SNCHO_2025}. The resulting chemical network consists of 103 species and 2067 reactions.  We discuss the reaction network in more detail in a companion paper \citep{yang2026cs2} and show that reaction rate uncertainties do not prevent accurate prediction of the CS$_2$ abundance.  For the stellar flux, we adopted that of GJ~876 from the \texttt{MUSCLES} survey III \citep[$T_{\rm eff}=3062$ K;][]{loyd_2016}, an M5-type star similar to TOI-6894 \citep[$T_{\rm eff}=3007$ K;][]{bryant_2025}, and scaled it to match the bolometric insolation of TOI-6894~b (5.54 $S_{\oplus}$; \citealt{bryant_2025}). For vertical mixing, we assumed a uniform eddy diffusion coefficient, $K_{\rm zz}$, of $10^8$ cm$^2$ s$^{-1}$. We note that the adopted atmospheric parameters for TOI-6894~b, $K_{\rm zz}$=$10^8$ cm$^2$ s$^{-1}$ and $T_{\rm int}$=100 K are representative of those commonly used in studies of Jupiter and other gas giant atmospheres \citep{BJORAKER1986579, li2012emitted}. We did vary $K_{\rm zz}$ from $10^6$ to $10^{10}$ cm$^2$ s$^{-1}$ to assess the sensitivity of the model to vertical mixing. Briefly speaking, we found that only \ce{NH3} (see the \ce{NH3} panel of Figure~\ref{fig:profiles}) and \ce{CO2} (not shown for brevity) were sensitive to variations in $T_{\rm int}$, whereas only \ce{CS2} showed a significant decrease with $K_{\rm zz}$, and only for values exceeding 10$^{10}$ cm$^2$ s$^{-1}$. The CH$_4$ abundance remains relatively insensitive to $T_{\rm int}$ up to 200 K. At $T_{\rm int}$  = 400 K, however, the deep atmospheric CH$_4$ abundance (at 100 bar) decreases by approximately two orders of magnitude, while the CH$_4$ abundance in the JWST-observable region decreases by approximately one order of magnitude. Despite this depletion, the CH$_4$ abundance in the upper atmosphere remains above 10$^{-4}$, which is sufficient to sustain the photochemical production of CS$_2$ in the atmosphere of TOI-6894~b.  Details of the sensitivity to atmospheric parameters, including $T_{\rm int}$, $K_{\rm zz}$, metallicity, and stellar UV activity, are presented in our companion modeling paper \citep{yang2026cs2}.

The converged 1D-photochemical models were used as inputs for transmission spectrum simulations with \texttt{PLATON} v6.2 \citep{Zhang_2019_platon}.  We computed opacities for \ce{CS2} using line lists from the latest \texttt{HITRAN} database \citep{GORDON_2026_Hitran}, including all isotopologues at their cosmic abundances.  We also experimented with using a preliminary ExoMol line list that has orders of magnitude more lines (private comm., Sergey Yurchenko), but found that it changes the spectrum negligibly due to the low temperature of our planet.

We perform these \texttt{PICASO} + \texttt{EPACRIS} + \texttt{PLATON} models for five atmospheric metallicities: 0.1, 1, 3, 10, and 30$\times Z_{\odot}$, where solar metallicity is based on \cite{lodders-2020}. 
 Figure~\ref{fig:spectra}b compares the resulting spectra for all metallicities with the data.  At 30$\times$, the model predicts too much CO$_2$ absorption (4.2--4.4 $\mu$m) as well as OCS (4.8--5.0 $\mu$m), because high metallicities favor more oxidized molecular compositions \citep{yang2024chemical}. It also predicts too little NH$_3$ absorption because higher metallicities favor N$_2$ over NH$_3$. At 0.1$\times$ solar, in contrast, the model predicts too little CO$_2$ and too much NH$_3$, for exactly the same reasons. The models which best explain the data, therefore, are the intermediate metallicities: 3$\times$ and 10$\times$ solar.

Figure \ref{fig:profiles} shows the abundance profiles predicted by the 3$\times$ solar model.  All major species have quench pressures deeper than 3 bar, which is in turn much deeper than the infrared photosphere. Photochemistry picks up in earnest at pressures below 10 mbar, increasing the mixing ratio of CS$_2$ to a maximum of $3 \times 10^{-5}$ by 0.2 mbar.  H$_2$S, the dominant carrier of S below 0.3 mbar, drops abruptly above that level because photochemistry converts almost all of it to CS$_2$.  NH$_3$, which equilibrium chemistry predicts should be the overwhelmingly dominant carrier of nitrogen at photospheric temperatures and pressures, instead only carries 14\% of the nitrogen because that is the equilibrium at the quench point.  This ratio decreases to 4\% at 10$\times$ solar because high metallicities favor N$_2$ over NH$_3$.

\subsection{Semi-free retrieval}
\label{subsec:semi-free}
Our RCP forward models assumed perfect knowledge of all parameters and perfect modelling of all the relevant physics.  This is, of course, not realistic.  We could go to the other extreme and allow all parameters to vary in a free retrieval, which we do in Appendix \ref{appendix:free_retrieval}, but this runs the risk of finding physically and chemically implausible solutions.  We instead opt for a happy medium, which we call the semi-free retrieval.

In the semi-free retrieval, we start with a RCP model and retrieve on \textit{perturbations} of its temperature profile and volume mixing ratio (VMR) profiles.  Our free parameters include $\Delta T$, which is a uniform temperature offset applied across all pressure levels of the T-P profile, and $\Delta \log X_i$, which is added to the log VMR of species i at all pressures.  In addition, we adopt as free parameters the planetary radius $R_p$ at 1~bar, an opaque cloud deck parametrized by its top pressure $\log_{10} P_{\rm cloud}$, and a haze parametrized by a scattering amplitude $\log_{10} a$ and a scattering slope $m$: $\sigma_h(\lambda) = \sigma_R (1\,\upmu \mathrm{m}) a (\lambda  / \upmu \mathrm{m})^{-m}$, where $\sigma_R$ is the Rayleigh opacity.  We also include an error inflation term, parameterized as a constant excess error added in quadrature to the photon noise at every wavelength bin.  By preserving the shapes of the temperature and log VMR profiles while letting their absolute values vary, we make some use of the physical wisdom of the RCP models while still giving the retrieval freedom to explore.  

\begin{figure*}[ht!] 
    \centering 
    \includegraphics[width=1.0\textwidth]{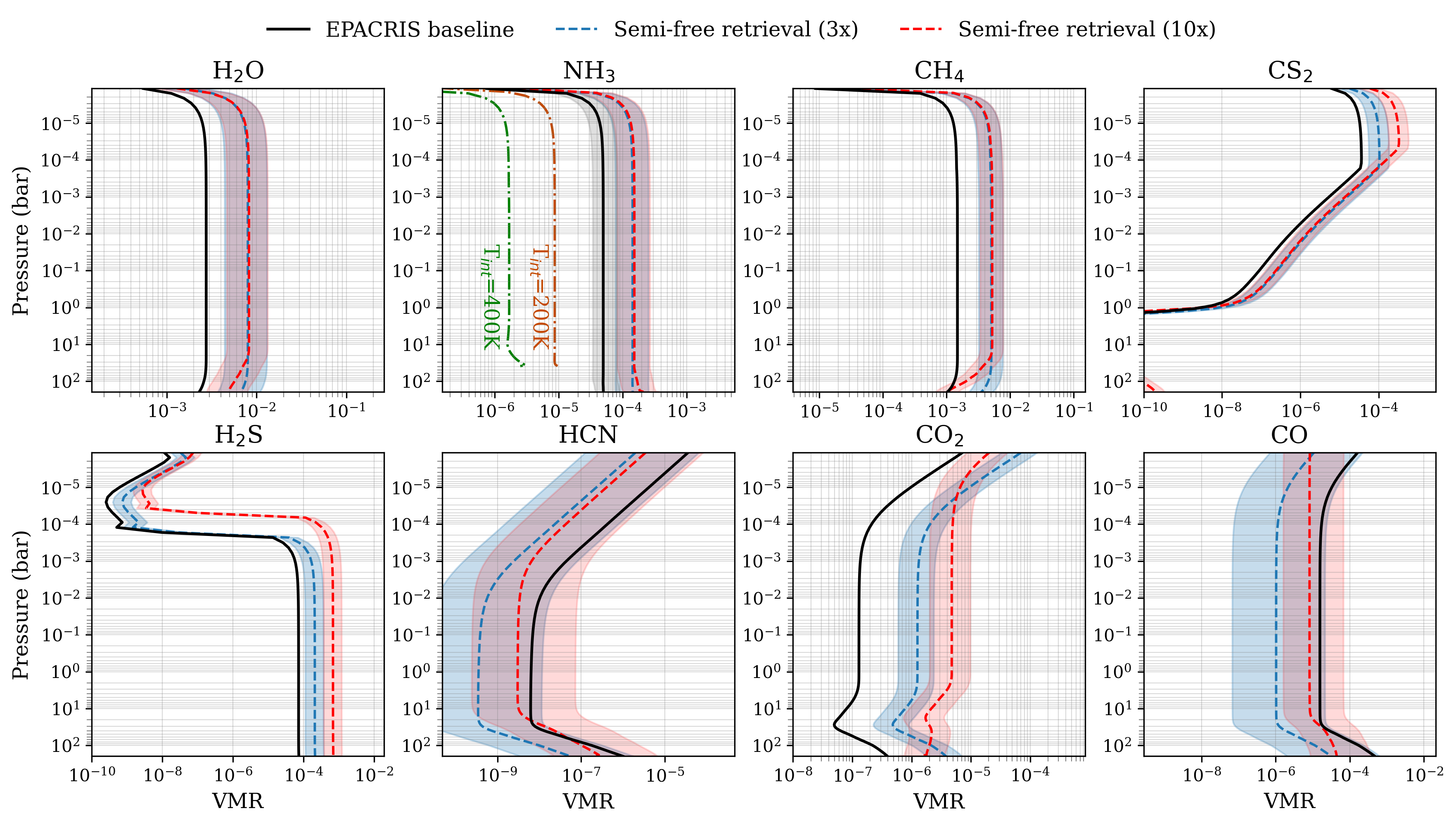}
    \caption{VMR profiles of the retrieved species. The solid black line represents the 3$\times$ solar metallicity photochemistry model from \texttt{EPACRIS}. The dashed line and shaded band show the median and 1$\sigma$ credible interval of the semi-free retrieval, in which each VMR profile is scaled by a multiplicative factor (with H$_2$S and \ce{CS2} sharing a common factor), blue for 3$\times$ metallicity baseline model and red for 10$\times$. We also show the 3$\sigma$ \ce{NH3} VMR constraint (gray shaded band), along with the profiles computed for different assumed values of $T_{\rm int}$ (the other species are not as sensitive to the internal temperature).
    }
    \label{fig:profiles} 
\end{figure*}

\begin{table}[]

\centering
\begin{tabular}{lccc}
\hline
\multicolumn{1}{l}{\textbf{Parameter}} & \textbf{Priors}          & \multicolumn{2}{c}{\textbf{Posteriors}}           \\ \hline
\textbf{Profiles}                      & \multicolumn{1}{l}{}     & \textbf{3x}             & \textbf{10x}            \\ \hline
$\log\Delta X_\mathrm{{CH_4}}$                             & $\mathcal{U}(-3,3)$ & $0.53^{+0.18}_{-0.20}$    & $0.09^{+0.17}_{-0.19}$    \\
$\log\Delta X_\mathrm{{CO_2}}$                            &      $\mathcal{U}(-3,3)$                    & $0.99^{+0.27}_{-0.33}$    &$-0.59^{+0.31}_{-0.37}$   \\
$\log\Delta X_\mathrm{{CO}}$                              &         $\mathcal{U}(-3,3)$                 & $-1.18^{+1.31}_{-1.19}$   & $-1.98^{+0.91}_{-0.71}$   \\
$\log\Delta X_\mathrm{{H_2O}}$                             &               $\mathcal{U}(-3,3)$           & $0.46^{+0.21}_{-0.26}$    & $-0.01^{+0.21}_{-0.25}$   \\
 $\log\Delta X_\mathrm{{HCN}}$                             &        $\mathcal{U}(-3,3)$                  & $-1.26^{+1.52}_{-1.20}$   & $-1.43^{+1.38}_{-1.10}$   \\
$\log\Delta X_\mathrm{{NH_3}}$                            &          $\mathcal{U}(-3,3)$                & $0.46^{+0.25}_{-0.27}$    & $0.46^{+0.24}_{-0.27}$    \\
$\log\Delta X_\mathrm{{H_2S,CS_2}}$                         &       $\mathcal{U}(-3,3)$                   & $0.05^{+0.17}_{-0.18}$   & $-0.18^{+0.15}_{-0.16}$   \\
$\Delta T$~[K]                                     & $\mathcal{U}(-200,200)$             & $26^{+22}_{-20}$          & $18^{+22}_{-20}$             \\
$R_p~[R_J]$                               & $\mathcal{U}(0.77,0.94)$         & $0.832^{+0.002}_{-0.002}$ & $0.831^{+0.002}_{-0.002}$ \\
$\log P_{cloudtop}$~[Pa]                         & $\mathcal{U}(1,8)$                   & $>$ 2.6   & $>$ 2.6  \\
log a                       & $\mathcal{U}(-1,6)$                & $2.77^{+0.20}_{-0.21}$   & $2.79^{+0.19}_{-0.20}$    \\
m                            & $\mathcal{U}(-3,7)$                 & $1.21^{+0.25}_{-0.20}$    & $1.29^{+0.28}_{-0.21}$   \\
excess error~[ppm]                         & $\mathcal{U}(0,1000)$                & $637^{+42}_{-43}$    & $636^{+44}_{-42}$   \\ \hline
\end{tabular}

\caption{Priors and posteriors from the semi-free retrievals.  The lower limit of the cloudtop pressure was calculated using the 1st percentile of the posterior.  $a$ is the scattering amplitude and $m$ is the scattering slope.}
\label{table:retrieval_params}
\end{table}

There is one final tweak we adopt to increase physical realism: we force CS$_2$ and H$_2$S to share the same $\Delta \log X_i$.  Since H$_2$S and \ce{CS2} are the dominant sulfur carriers in the deep and upper atmosphere, respectively, scaling both species by a shared factor is equivalent to varying the total atmospheric sulfur inventory while preserving the partitioning between the two carriers set by the photochemistry.  H$_2$S is in any case only weakly constrained by the data, as its absorption features within the PRISM bandpass are weak and heavily blended with those of H$_2$O and CH$_4$.

Starting with either the 3$\times$ solar or 10$\times$ solar RCP models from \S~\ref{subsec:self_consistent}, we adopt the priors in Table \ref{table:retrieval_params} and run \texttt{PyMultiNest} \citep{feroz_2008_multinest, buchner_2014_pymultinest} using 5000 live points. For both 3$\times$ solar and 10$\times$ solar models, the retrieved $\Delta T$ is consistent with 0, implying the \texttt{PICASO} T-P profile is accurate. The haze and cloud parameters suggest aerosol opacity with a slope weaker than that of Rayleigh scattering (indicative of large grains), but with no evidence of an opaque cloud deck.  The posteriors of sampled parameters are listed in Table~\ref{table:retrieval_params}. We show the inferred VMR profiles in Figure~\ref{fig:profiles} for the retrieval based on the 3$\times$ solar model.  The retrieved abundances of all species are consistent with the RCP model. NH$_3$, however, offers a handle on the planet's $T_{\rm{int}}$ \citep{soni_signature_2024}, from which we place an upper bound of $T_{\rm{int}} < 200$~K. 

\begin{figure*}[ht]
    \centering
    \includegraphics[width=1.0\textwidth]{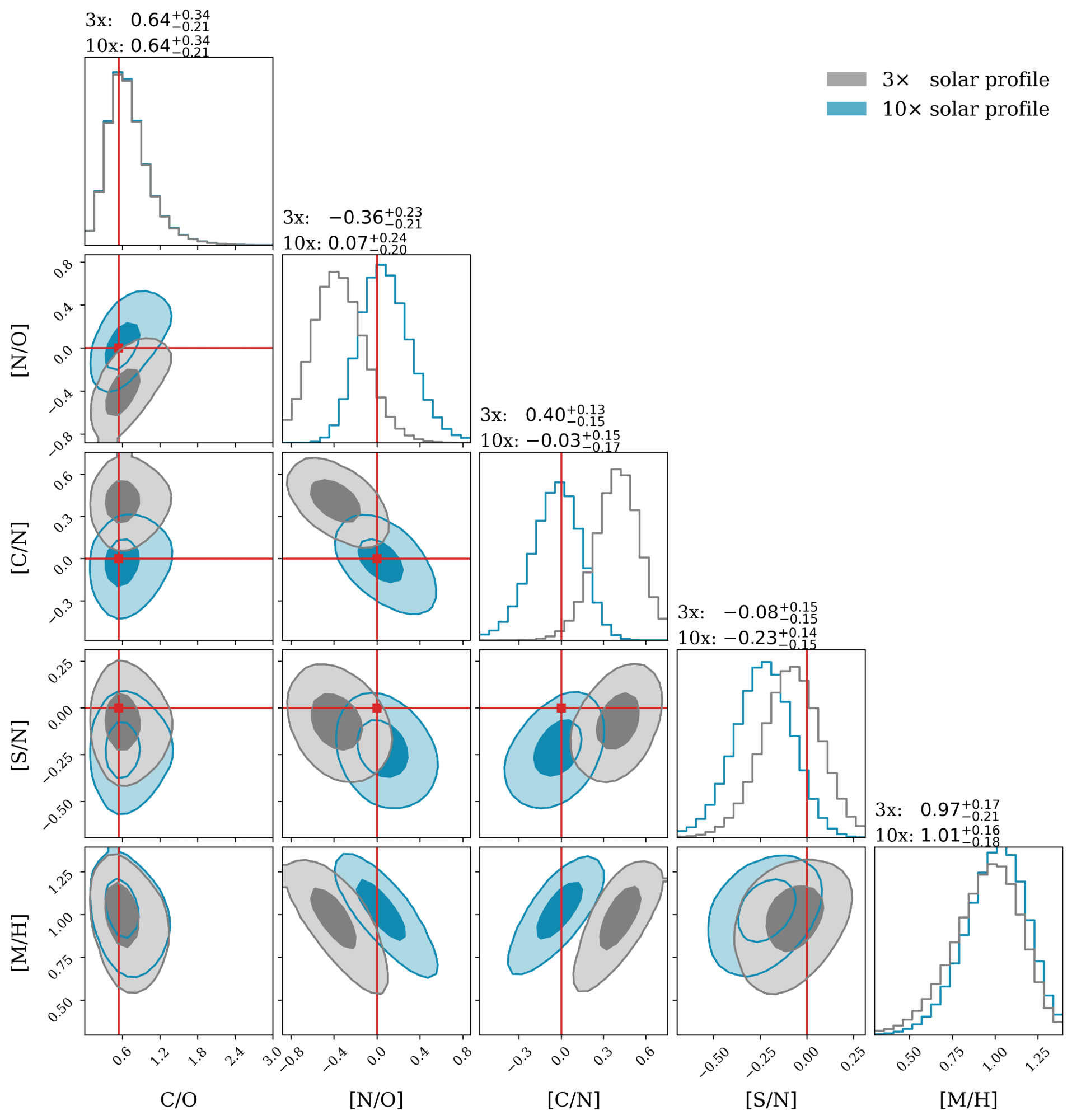}
    \caption{Derived elemental ratios from semi-free retrievals anchored on 3$\times$ (grey) and 10$\times$ solar metallicity (blue) PICASO + EPACRIS profiles. We express true C/O in linear scale, while N/O, C/N and S/N are given in log space normalized to solar values (i.e.  [X/H] = $\log_{10}\frac{\mathrm{(X/H)}}{\mathrm{(X/H)_\odot}}$). The red lines mark 1$\times$ solar abundance for reference. We find that these ratios are all consistent with solar values.}
    \label{fig:corner} 
\end{figure*}

On the basis of the molecular abundances inferred from the semi-free retrievals, we compute metallicity, C/O, N/O, C/N, and S/N ratios, which we show in Figure \ref{fig:corner}.  These are all consistent with 3--10$\times$ solar metallicity and solar abundance ratios. TOI-6894~b is the first transiting exoplanet for which any nitrogen species is visually evident in the spectrum. Nitrogen has been proposed as a tracer of planet formation \citep{ohno_nitrogen_2023, ohno_nitrogen_2023-1, turrini_tracing_2021}, but most of the nitrogen in TOI 6894b's photosphere resides in spectroscopically inactive N$_2$.  We can, however, report a 3$\sigma$ lower limit on the nitrogen abundance: [N/H]$>0.38$ from the retrieval based on the 3$\times$ solar model.

\subsection{1D-RCPE grid retrieval}
\label{subsec:schimera}
The final modelling approach we adopt is that of the grid retrieval.  Here, we perform a retrieval by interpolating from a grid of fully self-consistent \texttt{ScCHIMERA} + \texttt{Photochem} models \citet{bell_2023, welbanks_2024, Wiser_2025}.  By using different codes and a different modelling philosophy, we can cross-check the conclusions of the previous two subsections.

 The {\tt ScCHIMERA} RCE solver \citep{Piskorz_2018, Arcangeli_2018, Mansfield_2021} determines the radiative-convective equilibrium thermal structure under the assumption of thermochemical equilibrium.  Upon convergence, the temperature profile and equilibrium chemical abundances are fed to the photochemical kinetics solver {\tt photochem} \citep{wogan2025photochem} with VULCAN's \texttt{SNCHO\_photo\_network\_2025.txt} \citep{VULCAN_SNCHO_2025} chemical network. The resultant kinetically derived gas mixing ratios are then fed back in and fixed inside the {\tt ScCHIMERA} RCE solver to derive an updated temperature profile. This process is repeated 3 times, which is sufficient to ensure convergence.

We interpolate the {\tt Phoenix} stellar atmosphere spectral library \citep{husser2013new} at $T_{\rm{eff}}$ = 3007\,K, $log g$ = 5.0 to obtain the input top-of-the atmosphere stellar flux.  Similar to \texttt{EPACRIS} above, in {\tt photochem} we use the \texttt{MUSCLES} UV template spectrum. The 1D-RCPE grid is generated as a function of [M/H] (0 - 2.0 in steps of 0.125 dex), C/O (0.1 - 0.8 in steps of 0.1), and irradiation temperature (375 - 450\,K in steps of 25\,K).  The vertical mixing, $K_{\rm zz}$, is fixed to a constant 1$\times$10$^8$ cm$^2$\,s$^{-1}$ and the internal temperature to 150\,K. This results in 640 1D-RCPE model atmospheres.

We then use the {\tt PyMultiNest} nested sampling tool \citep{feroz_2008_multinest,buchner_2014_pymultinest} to fit the transmission spectrum and obtain constraints on the composition. We fit for the three grid parameters-- [M/H], C/O, and T$_{\rm{irr}}$--as well as several nuisance parameters that include a vertically uniform gray cloud opacity, a power law haze slope and amplitude \citep{des_etangs_2008}, a two-terminator patchy cloud fraction \citep{line_2016}, 1 bar radius scaling,  and stellar heterogeneity \citep{rackham_2018} assuming spots (spot covering fraction and temperature) for a total of 10 free parameters.  A description of the transmission spectrum routine and opacity sources can be found in \citet{bell_2023, welbanks_2024}. One notable change is the inclusion of the CS$_2$ opacity with the updated {\tt HITRAN} line list described above. 

We show the best fitting model in Figure \ref{fig:spectra}.  Our grid retrieval gives [M/H]=$0.46 \pm 0.08$ ($2.9 \pm 0.6$ times solar metallicity), consistent with the conclusions from the PICASO + EPACRIS forward model and from our semi-free retrieval.  It implies a C/O ratio of $0.69 \pm 0.06$, consistent with both the solar value and with our semi-free retrieval.  We find a haze with much stronger extinction than molecular Rayleigh scattering (log(a) = $2.71 \pm 0.13$ at a reference wavelength of 1 $\mu$m), but with much weaker slope ($\beta = 0.49 \pm 0.09$).  We also find no evidence of stellar heterogeneity, with an upper limit on the spot fraction of $\sim$0.1.  These haze and stellar heterogeneity inferences are consistent with the semi-free retrieval.


\subsection{Interior modelling}
\label{subsec:interior_modelling}
Atmospheric retrieval results in hand, we use the \texttt{SMILE} internal structure model \citep{nixon_2021} to constrain the heavy element mass $M_z$ of the planet in order to determine how surprising the planet's existence around a late M dwarf ought to be. \texttt{SMILE} calculates the radius of a planet for a given mass and composition by solving the equations of hydrostatic equilibrium and mass conservation. For this study, the planet consists of a solid ice+rock nucleus with an ice:rock ratio of 1:1, and a gaseous envelope consisting of H, He and H$_2$O. Following \citet{nixon_2024}, the proportions of H/He and H$_2$O in the envelope are chosen so that the atmospheric mean molecular weight is equivalent to that of an atmosphere with 12.6 $\times$ solar metallicity, on the upper end of our retrieval results.  In other words, H$_2$O is used as a proxy for all metals in the atmosphere. For both the nucleus and the envelope, we construct a mixed EOS using individual EOS tables for each component, combined using the linear mixing approximation. The heavy element mass $M_z$ is the sum of the mass of the nucleus and the mass of H$_2$O in the envelope.

We compute a grid of models spanning a range of values of the total mass and the heavy element mass of the planet. We subsequently explore the subset of models whose mass and radius are consistent with the measured values for TOI-6894~b to $1\sigma$. We obtain $M_z = 11.1 \pm 1.7 \, M_{\oplus}$, with approximately 70\% of the heavy element mass located in the solid nucleus, and the remainder in the envelope. This inferred heavy element mass is fully consistent with the heavy element mass of $12 \pm 2 \, M_{\oplus}$ reported by \citet{bryant_2025}, which they derived assuming a solar metallicity atmosphere, indicating that this inference is insensitive to the precise atmospheric metallicity. We note, however, that our models assume a fixed $T_{\rm int}=100\,$K, following the approach used for the RCP forward models. Exploring a wider range of values of $T_{\rm int}$ could lead to a larger uncertainty on the inferred metallicity and core mass fraction.


\section{Discussion}
\label{sec:discussion}
TOI-6894 is the lowest mass star known to host a transiting giant planet.  Considering the severe stellar contamination that accompanied the three M dwarf giant planet transmission spectra published so far --- those of HATS-75~b \citep{ashtari_2026}, TOI-5293Ab \citep{kanodia_2026}, and TOI-5205~b \citep{canas_2026} --- the cleanliness of our light curve and spectra is remarkable. This is partly because M dwarfs have a range of activity levels and TOI-6894 happens to be very quiet, with no detectable rotation period or flares in the TESS light curve \citep{bryant_2025}. Also, TOI-6894~b has low gravitational potential, which matters because the size of TLS-induced pseudofeatures is proportional to the transit depth $(R_p/R_s)^2$ (Eq 1 of \citealt{rackham_2018}) while the size of a one scale height atmospheric feature is $\frac{2\pi R_p H}{R_s^2}$.  The ratio of the former to the latter is proportional to $\frac{\mu}{T} \frac{GM_p}{R_p}$--TLS is less prominent at lower gravitational potential.

The other striking fact about TOI-6894~b, especially in contrast to the low metallicities and supersolar C/O ratios tentatively claimed for the three previously observed giant planets orbiting M dwarfs, is its surprisingly unsurprising composition: a metallicity of a few to several times solar, with C/O, N/O, and S/N ratios consistent with solar values.  This is very similar to the composition of Jupiter \citep{rensen_2023}, Saturn \citep{briggs_1989,fletcher_2009}, and the four HR 8799 planets \citep{xuan_2026}.  It is quite possibly similar for many other gas giants, but no other transiting exoplanet has a measured elemental inventory for all of C, N, O, and S.  A rare combination of properties made this possible for TOI-6894~b: the small star, the low gravity, the temperate conditions, and the photochemical conversion of the sulfur inventory into CS$_2$.  The planet is warm enough that none of its major C-, O-, N-, or S-bearing reservoirs are sequestered into condensates, but cool enough that its volatiles all form species with strong spectral features observable by JWST.


Our interior models show that even at 12$\times$ solar, the metallicity of TOI-6894~b is low enough that it does not significantly affect the bulk metallicity.  The bulk metal content of the planet would still be $\sim12 M_\oplus$, amounting to $\sim$23\% of the total mass.  For comparison, the bulk metallicity is $0.21 \pm 0.01$ for Saturn, which is 80\% more massive than TOI 6894b (c.f. \citealt{guillot_2023}, and references therein).  The Saturn-like bulk and atmospheric metallicities, along with the Sun-like abundance ratios, suggest that even M5 dwarfs can form 50 $M_\oplus$ giants through the standard mechanism of core accretion in a protoplanetary disk.  12 $M_\oplus$ is 0.9\% of the stellar mass in metals, since the stellar metal mass fraction is $0.0189 \pm 0.0037$ \citep{bryant_2025}. One can obtain this much metal by invoking a high disk-to-star mass ratio of $\sim10$\% and a disk-to-planet conversion efficiency $\sim$9\%, which are both high but not clearly unreasonable \citep{lin_2018,mercer_2020}.  If the unusual compositions of the other three GEMS are validated, they would be suggestive of a different formation channel for hotter and more massive giants orbiting M dwarfs.  It is interesting that in terms of the ratio between planetary and stellar metal mass, TOI-6894~b is in fact more extreme than HATS-75~b (0.3\%) and comparable to TOI-5205~b (0.5--1.0\%), a consequence of the lower stellar mass and lower stellar metallicity.

Our observations are in theory accurate enough to constrain formation mechanisms through abundance ratios.  For example, \cite{turrini_tracing_2021} show that gas envelope C/N, C/O, N/O, and S/N ratios can differ by a factor of $\sim$2 from the stellar ratio at different formation locations if planet metallicity is planetesimal accretion dominated, and by a factor of several if metallicity is gas accretion dominated.  If their simulation were fully applicable to TOI 6894b--which it is not, since it simulates a 1 $M_J$ planet orbiting a solar analogue with a final orbit at 0.4 AU--our observations would be precise enough to rule out the combination of gas-accretion-dominated metallicity and large formation semi-major axis ($>10$ AU).  We encourage planet-specific modelling to constrain formation scenarios from our observational results.  In the future, it is conceivable that the error bars on these elemental ratios can be reduced much farther if fully self-consistent RCPE models can be made fast enough to run in a retrieval, and accurate enough that model error is negligible.

Before concluding, we point out one important direction for future work: improving the agreement between data and model.  As Figure \ref{fig:spectra} shows, no model we have tried can match the low transit depths at the red end.  Our models also underpredict the methane peak at 1.8 $\mu$m while overpredicting the blue edge of the methane peak at 1.4 $\mu$m.  Plausible explanations for these mismatches include missing opacities, aerosol absorption features, and stellar spectral features (imprinted on the spectrum by TLS), but we have so far not found a conclusive resolution.


\section{Conclusion}
\label{sec:conclusion}
Among all the known transiting GEMS, TOI-6894~b is the coldest and orbits the lowest mass star.  Among all the GEMS with JWST observations, TOI-6894~b stands out for its exceptionally well-behaved star, high transit spectroscopy metric, and large inventory of observable molecules, leading us to infer a slightly super-solar metallicity and solar-like C/O, N/O, and S/N.

TOI-6894~b has photochemically produced CS$_2$, like a growing number of exoplanets: TOI-270d \citep{holmberg_possible_2024}, WASP-80b \citep{triantafillides_identification_2026}, and V1298 Tau e \citep{dai_photochemical_2026}.  The importance of CS$_2$ in carrying the sulfur inventory is demonstrated both by our own models and by previous work (e.g. \citealt{mukherjee_2025}).  In a companion paper \citep{yang2026cs2}, we demonstrate that our new, automatically generated photochemical reaction network can accurately predict the amount of CS$_2$ on this planet, and that CS$_2$ is expected to be common on gas giants of $T_{\rm eq}=400-800$ K.

Much more science is possible with our dataset.  Throughout the paper, we have been assuming that TOI-6894~b has homogeneous limbs due to its low temperature and subsequently high radiative timescales.  In a follow-up paper (Q.\ Xue et al. in prep), we will demonstrate that although the differences between the morning and evening limb are much smaller than for some previously observed warm and hot gas giants, they are clearly visible.

TOI-6894~b is a spectacular planet, an oddball even among the exotic GEMS.  We encourage future observational and theoretical efforts to understand this planet, and to discover other planets like it.  In particular, the quietness of the star provides hope that atmospheric characterization of exoplanets orbiting late M dwarfs --- from the gas giants to the habitable zone rocky worlds --- is possible.

\begin{acknowledgments}
This work is based on observations made with the NASA/ESA/CSA James Webb Space Telescope. The data were obtained from the Mikulski Archive for Space Telescopes at the Space Telescope Science Institute, which is operated by the Association of Universities for Research in Astronomy, Inc., under NASA contract NAS 5-03127 for JWST. These observations are associated with program \#GO-8696. Support for this program was provided by NASA through a grant from the Space Telescope Science Institute.

We thank Tyler Fairnington for his help with adding semi-free retrieval capability to \texttt{PLATON}. M.Z. thanks the Heising-Simons Foundation for his 51 Pegasi b fellowship. C.P.-G. acknowledges support from the E. Margaret Burbidge Prize Postdoctoral Fellowship from the Brinson Foundation, and from the Suzuki Postdoctoral Fellowship. A.J. acknowledges support from Fondecyt project 1251439.
\end{acknowledgments}

\begin{contribution}
MZ proposed the observations, led the team, supervised QX, and wrote 1/3 of the paper. QX performed the \texttt{SPARTA} data reduction and the semi-free retrieval, in addition to writing half of the paper. JY produced the new photochemical reaction network and ran the  \texttt{EPACRIS} + \texttt{PLATON} analysis. VN ran the \texttt{PICASO} models. MRL performed the grid retrieval. GF performed the \texttt{Tswift} data reduction. MN performed the interior structure modeling. JLB and EMRK provided scientific leadership and assisted with the paper writing. EB, DB, VVE, AJ, ES, and AHMJT led the discovery and mass measurement of TOI-6894~b. PG, LW, MB, JMD, JJF, VP, CPG, and KBS participated in the original proposal and contributed to discussions.

\end{contribution}

%
\facilities{JWST(NIRSpec)}

\software{\texttt{astropy} \citep{astropy:2013,astropy:2018,astropy:2022}, \texttt{numpy} \citep{harris2020array}, \texttt{scipy} \citep{2020SciPy-NMeth}, \texttt{matplotlib} \citep{Hunter:2007}, \texttt{SPARTA} \citep{kempton_reflective_2023}, \texttt{PLATON} \citep{Zhang_2019_platon, Zhang_2025_platon}, \texttt{PyMultiNest} \citep{feroz_2008_multinest}, \texttt{jwst} \citep{bushouse_2025_16280965}, \texttt{ExoTiC-LD} \citep{husser2013new,Grant2024}, \texttt{ldtk} \citep{Parviainen2015}, \texttt{Tswift} \citep{kirk_jwstnircam_2024}, \texttt{batman} \citep{kreidberg_batman_2015}, \texttt{emcee} \citep{foreman-mackeyEmceeMCMCHammer2013}, \texttt{EPACRIS} \citep{yang_2024}, \texttt{PICASO} \citep{mang2026picaso}}


\appendix

\section{Data Reduction and Analysis} \label{appendix:data}
\subsection{\texttt{SPARTA}} \label{appendix:sparta}
The uncalibrated \texttt{\_uncal.fits} were reduced with \texttt{SPARTA} \citep{kempton_reflective_2023}, an open-source pipeline developed specifically for JWST time-series observations of transiting exoplanets. \texttt{SPARTA} implements an end-to-end reduction from the raw \texttt{\_uncal.fits} data products through to spectroscopic light curves, performing detector-level processing, spectral extraction, and light curve fitting within a single lightweight framework rather than relying on the \texttt{jwst} calibration pipeline for the initial stages. We used the version of \texttt{SPARTA} that was described in \citet{Zhang_2024} and \citet{xue_jwst_2025}, with the NIRSpec\/PRISM extensions presented in Q.\ Xue et al.\ (in prep.). 

Starting from the raw group-level ramps, \texttt{SPARTA} performs the detector-level calibration steps, including superbias correction, reference pixel correction, nonlinearity correction, dark subtraction, cosmic-ray rejection, and up-the-ramp fitting, following the standard JWST pipeline recommendations, but implemented independently of \texttt{jwst}. After the nonlinearity correction step, \texttt{SPARTA} also applies a column-wise 1/$f$ correction at the group level. For each column, the median value of the region that is 8 pixels above and below the spectral trace is subtracted from that column. Following up-the-ramp slope fitting, \texttt{SPARTA} performs a second background-subtraction step on the resulting \texttt{\_rateint} images. For each integration and each column, the median flux in off-trace background regions located 10 pixels above and below the spectral trace is subtracted from every pixel in that column.

\begin{figure*}[ht!] 
    \centering 
    \includegraphics[width=0.7\textwidth]{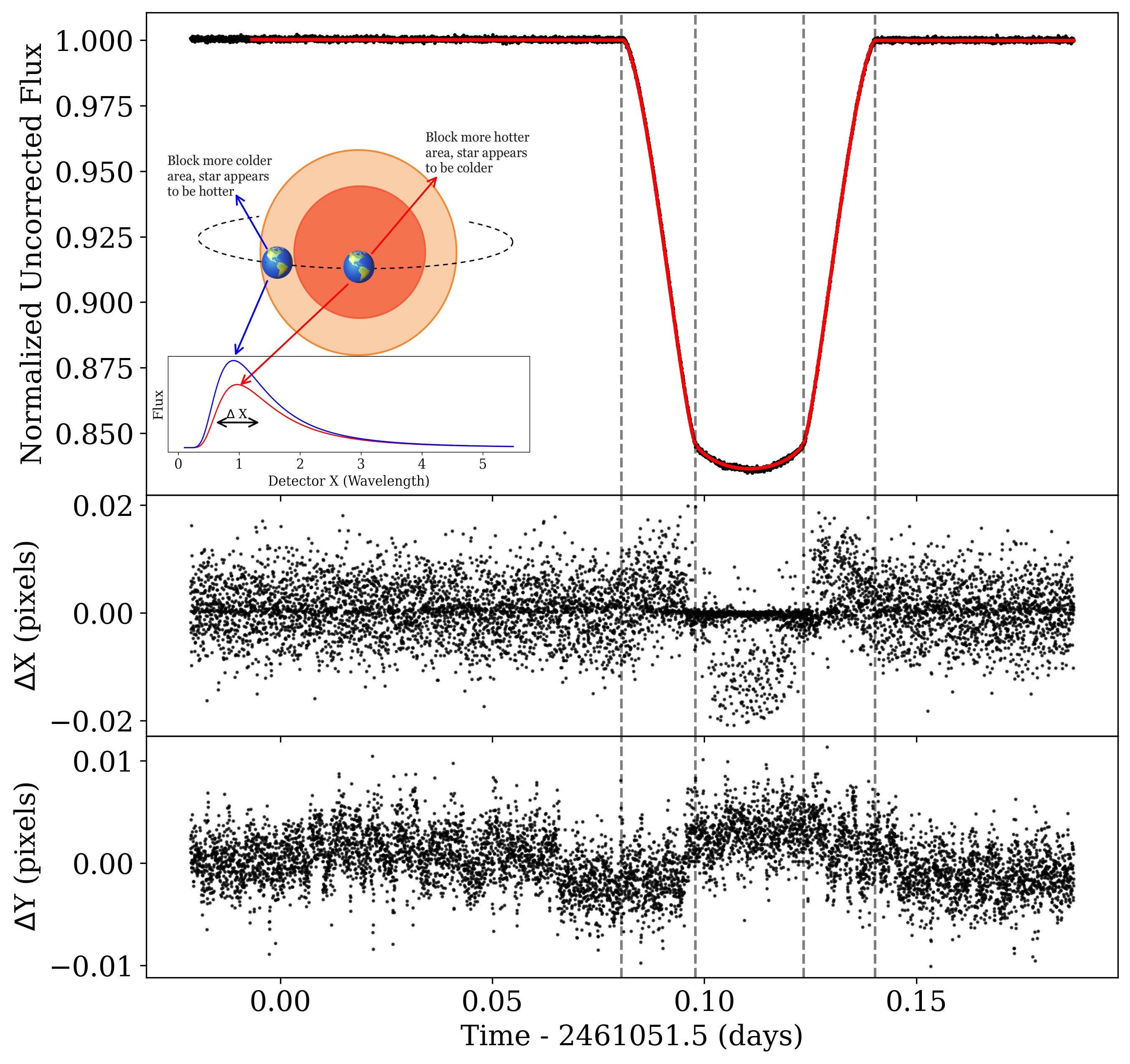} 
    \caption{Spectral centroids offsets in the detector x and y direction throughout the observation. The offsets track the shape of the light curve and are approximately symmetric about mid-transit. This is due to the large transit depth causing wavelength-dependent shifts in the stellar spectra as the planet traverses the stellar disc (see illustration in the top panel). These centroid variations are therefore excluded as decorrelation vectors in the lightcurve fits.\label{appendix:figure_lightcurve}}
\end{figure*}

To track sub-pixel pointing drift across the time series, we registered each integration against a high signal-to-noise median template that was constructed as the median of all time-series rate images, with bad pixels and 15$\sigma$ outliers replaced by linear interpolation across each detector row. We find, however, that the pixel offsets vary drastically across the transit and are strongly correlated with the flux variation. The amplitudes and shapes of $\Delta$x and $\Delta$y track the light curve and are approximately symmetric about mid-transit. This is due to the very large transit depth of TOI-6894~b ($\sim$15\%). Near ingress and egress, the planet blocks the cooler limb, so the disc-integrated stellar spectrum appears hotter and is shifted bluewards, whereas near mid-transit the planet blocks the hotter disc centre, so the spectrum appears cooler and shifts redwards (see Figure~\ref{appendix:figure_lightcurve}). Given these artefacts, we exclude the fitted offsets as decorrelation vectors in the spectroscopic lightcurve fits.

The spectrum was then extracted using optimal extraction \citep{horne_optimal_1986}. Details of \texttt{SPARTA} implementation can be found in \citet{kempton_reflective_2023}. We obtained the latest PRISM wavelength solution from the x1d products generated by STScI. We tested extraction half-widths from 4 to 15 pixels in steps of 1 pixel, and adopted a half-width of 5 pixels, which minimised the median absolute deviation of the white lightcurve residuals.

\subsubsection{Light curve fitting}
In the 2D $F(t,\lambda)$ grid we obtain, we manually remove the column at 1.6009 $\mu$m and reject integrations that have flux values deviating from a running median by more than 4$\sigma$. 

We fit the lightcurves combining the \texttt{batman} \citep{kreidberg_batman_2015} transit model and a linear-in-time systematics model $$F_{sys}(t) = c_0 + c_1(t - \bar{t}),$$ where $\bar{t}$ is the mean of the observation time. We trimmed the first 500 integrations to get rid of any substantial detector settling effect. A combined exponential-ramp multiplied with linear-in-time model $F_{sys}(t) = Ae^{(-\frac{t-t_0}{\tau})} + m(t-\bar{t}) $ and a quadratic-in-time model $F_{sys}(t) = c_0 + c_1(t - \bar{t}) + c_2 (t-\bar{t})^2$ yielded higher Bayesian Information Criterion (BIC) values in all wavelength channels compared to the simpler linear-in-time model, indicating no statistical support for the additional parameters. The orbital period was fixed to P=3.37077~d from \citet{bryant_2025}.

For the white light curve, the free parameters were $R_p/R_s$, mid-transit time $T_0$, scaled semi-major axis $a/R_s$, orbital inclination $i$, the two systematics coefficients ($c_0$, $c_1$) and the two power-2 limb-darkening coefficients (c, $\alpha$). We report the best fit values from the 0.8 -- 5.2 $\mu$m white lightcurve to be T$_0 = 2461051.610618 \pm 1.4\times10^{-6}$~d, $a/R_\star = 24.857^{+0.014}_{-0.011}$, $i = 89.614\pm{0.003}$~deg and $R_p/R_\star = 0.3889\pm0.0002$.

\subsubsection{Limb darkening}
The choice of limb-darkening parameterization has been shown to be a non-negligible source of systematic error in high-precision light curves \citep{coulombe_biases_2024}. Moreover, TOI-6894 is a cool M host whose intensity profile is poorly described by the parameterizations developed for Sun-like stars. We therefore treat limb darkening carefully and describe our modeling approach below.

The quadratic law,
\begin{equation}
\label{eq:quad}
\frac{I(\mu)}{I(1)} = 1 - u_1\,(1 - \mu) - u_2\,(1 - \mu)^2,
\end{equation}

is the most widely used limb-darkening parametrization in transit light curve analyses. We sampled its coefficients using the Kipping reparametrization \citep[$q_1, q_2$;][]{kipping_efficient_2013}. We first left both coefficients free in every pixel-level spectroscopic channel. We found that the coefficients become essentially unconstrained in the low-S/N channels, especially redward of 5 $\mu$m.
 
We next fixed the coefficients to theoretical values computed from PHOENIX model atmospheres with \texttt{ExoTiC-LD} \citep{husser2013new,Grant2024} using T$_{\rm{eff}} = 3007$~K, $\log g$ = 5.0 and [M/H] = 0.14 \citep{bryant_2025}. Of the various minimum-$\mu$ cuts applied when fitting the quadratic law to the model intensity profile, $\mu_{min}$=0.5 gave the best overall agreement with the freely-sampled $(q_1, q_2)$; lower cuts, which include more of the poorly-modelled stellar limb, made the agreement worse.
 
Finally, we tested a hybrid approach in which the PHOENIX model predictions were rescaled to match empirically-fitted coefficients. We fitted $(q_1, q_2)$ on high-S/N wavelength bins ($\Delta\lambda=0.5~\mu$m per bin), constructed by binning many native channels together, and used these well-constrained values to rescale the PHOENIX-predicted $(q_1, q_2)$ at the native channel resolution.
 
Both the fixed-to-model and the hybrid approaches, however, produced anomalous residuals in a subset of wavelength channels, indicating that the imposed coefficients were inconsistent
with the data in those channels. Moreover, the BIC was \emph{higher} for the fixed-coefficient fits (i.e., disfavored) relative to the free fits, despite the smaller number of free parameters. 

We repeated the same exercises with the power-2 law,
\begin{equation}
\label{eq:p2}
\frac{I(\mu)}{I(1)} = 1 - c\,\bigl(1 - \mu^{\alpha}\bigr),
\end{equation}
which captures the steep centre-to-limb intensity drop more accurately than the quadratic law \citep{morello_2017}. Since the power-2 parameterization is not available in \texttt{ExoTiC-LD}, we computed the power-2 coefficients with \texttt{ldtk} \citep{Parviainen2015} by fitting Equation~\ref{eq:p2} directly to the PHOENIX-based model intensity profile in each channel. This has another advantage that \texttt{ldtk} evaluates the intensity profile on an arbitrarily fine $\mu$ grid before the law is fitted, giving us direct control over the $\mu$ sampling and range.
 
As with the quadratic law, leaving $(c, \alpha)$ free produced well-constrained coefficients at high S/N but poorly-constrained coefficients at low S/N, while both fixing the coefficients to the \texttt{ldtk} predictions and anchoring them to high-S/N empirical values introduced similar anomalous residuals at the limb in some channels and a disfavoured BIC relative to the free fits.

For the reasons above, we leave the limb darkening parameters free in our final analysis. While the BIC comparison between the power-2 and quadratic laws is inconclusive at pixel level, the white-light curve gives $\Delta\mathrm{BIC} = -13$ in favour of the power-2 law. We therefore adopt the power-2 law for the fiducial transmission spectrum.

\subsection{\texttt{Tswift}} \label{appendix:Tswift}

Tswift is an independent JWST data reduction pipeline developed by Guangwei Fu.  In our Tswift reduction, we began from the uncalibrated (\texttt{uncal}) products and processed each segment through a custom implementation of the \texttt{jwst} Stage-1 (\texttt{calwebb\_detector1}) pipeline. We applied the group-scale, data-quality initialization, saturation flagging, superbias, reference-pixel, linearity, and dark-current steps with the default reference files. Before ramp fitting, we performed a group-level background subtraction to remove the $1/f$ striping noise. For every group of every integration we computed the median of the off-trace background rows (the 7 topmost and 9 bottommost rows of the subarray) column-by-column and subtracted it from that group. We then ran the jump-detection step (rejection threshold of 8\,$\sigma$) and fit the ramps to produce per-integration count-rate (\texttt{rateints}) images. Mid-integration timestamps were taken from the \texttt{INT\_TIMES} extension in BJD$_{\mathrm{TDB}}$.

Next, we applied a temporal outlier rejection by constructing the median image across all integrations and replacing any pixel deviating by more than 1000\,DN\,s$^{-1}$ from the median with the corresponding median value. We extracted the stellar spectrum by summing a fixed five-pixel-wide aperture centered on the spectral trace, producing a one-dimensional stellar spectrum for each integration. Residual outlier points in the resulting spectroscopic light curves were removed with a moving-median filter applied column-by-column (window of 20 integrations, 8$\times$\,median-absolute-deviation threshold), yielding the cleaned set of per-column light curves used for the analysis.

We first fit the white-light curve, formed by summing detector columns. The transit was modeled with \texttt{batman} \citep{kreidberg_batman_2015} assuming a circular orbit and a quadratic limb-darkening law, multiplied by a linear-in-time baseline. We held the orbital period fixed at $P=3.370772$\,d and sampled the posterior with \texttt{emcee} \citep{foreman-mackeyEmceeMCMCHammer2013}, using 32 walkers and 5000 steps with the first 1000 discarded as burn-in. The free parameters were the planet-to-star radius ratio, the scaled semi-major axis $a/R_\star$, the inclination, the transit time, both limb-darkening coefficients, the baseline slope, and a normalization constant. This yielded $R_p/R_\star = 0.3897 \pm 0.0002$, $a/R_\star = 24.897^{+0.012}_{-0.013}$, and $i = 89.612^{+0.004}_{-0.004}\ \mathrm{deg}$.

To construct the transmission spectrum, we fit each detector column independently using \texttt{batman} and \texttt{scipy} non-linear least squares. We fixed $a/R_\star$, the inclination, the transit time, and the period to the white-light values, and floated only the radius ratio, the second quadratic limb-darkening coefficient, a linear baseline slope, and a normalization constant. The first limb-darkening coefficient of each channel was fixed to the value predicted by the PHOENIX stellar models via \texttt{ExoTiC-LD} \citep{Grant2024}, computed for the host parameters
$T_{\rm eff}=3007$\,K, $\log g = 5.0$, and $[\mathrm{M/H}]=0.14$. Per-channel transit depths were taken as $(R_p/R_\star)^2$, with uncertainties propagated from the fit covariance matrix scaled by the out-of-transit scatter of each light curve. The wavelength solution was adopted from the STScI Stage-2 pipeline extracted-spectrum (\texttt{x1dints}) product and mapped onto the detector columns.

\section{Free retrieval}
\label{appendix:free_retrieval}

We also performed a fully free retrieval in which a vertically-constant volume mixing ratio was retrieved for each molecular species. The retrieved species were H$_2$O, NH$_3$, CH$_4$, HCN, CO$_2$, CO, H$_2$S, and CS$_2$, alongside the planetary radius $R_p$ at a reference pressure of $1$\,bar, an isothermal temperature $T$, a grey cloud top pressure $\log_{10} P_{\rm cloud}$ and haze scattering parameters. The priors given for the VMRs are log10 uniform from -10 to -1, and uniform from 200 to 600 K for the temperature. For the other parameters, we use the same priors as listed in Table~\ref{table:retrieval_params}. We reported the median and $\pm 1\sigma$ values of the retrieved parameters to be $R_p/R_J = 0.836\pm0.001$, $T_p=447^{+18}_{-16}$~K, log$_{10}$ scatter factor=$3.04\pm0.16$, scatter slope=$1.74^{+0.55}_{-0.38}$, log$_{10}P_\mathrm{cloudtop}$~[Pa]$>2.5$, $\log X_{\mathrm{H_2O}}=-1.40^{+0.12}_{-0.15}$, $\log X_{\mathrm{NH_3}}=-3.19^{+0.24}_{-0.26}$, $\log X_{\mathrm{CH_4}}=-1.67^{+0.16}_{-0.16}$, $\log X_{\mathrm{HCN}}=-5.85^{+1.82}_{-2.76}$, $\log X_{\mathrm{CO_2}}=-4.18^{+0.41}_{-0.48}$, $\log X_{\mathrm{CO}}=-6.62^{+2.12}_{-2.25}$, $\log X_{\mathrm{H_2S}}=-3.38^{+0.96}_{-4.07}$, $\log X_{\mathrm{CS_2}}=-4.06^{+0.21}_{-0.21}$.


We found, however, that treating H$_2$S as an independent free parameter drives its retrieved abundance to unphysically high values, corresponding to an inferred S/H ratio of $\sim200\times$ solar. To avoid this runaway, we performed an additional retrieval in which the total sulfur abundance was partitioned between H$_2$S and CS$_2$ as described below.

In the sulfur partition free retrieval, we replaced the independent H$_2$S and CS$_2$ free parameters with a single log-uniform parameter $\log_{10} f_{\rm S,tot}$ representing the total atomic sulfur volume mixing ratio in the atmosphere. We assigned equal molecular abundances to the two sulfur carriers, $f_{\rm H_2S} = f_{\rm CS_2} = f_{\rm S,tot}/3$, which partitions the atomic sulfur as $1/3$ in H$_2$S and $2/3$ in CS$_2$, the latter contributing two sulfur atoms per molecule. The prior for $\log_{10} f_{\rm S,tot}$ is given as uniform from -10 to -1. All other molecules together with the $R_p$, $T$, $\log_{10} P_{\rm cloud}$, and haze parameters were retrieved as in the free retrieval above. By construction, this setup enforces conservation of the atomic sulfur budget and prevents the H$_2$S runaway, while still allowing both species to contribute to the spectrum. For this free-retrieval, we report $R_p/R_J = 0.836\pm0.001$, $T_p=447\pm18$~K, $\log_{10}$ scatter factor=$3.04\pm0.17$, scatter slope=$1.74^{+0.59}_{-0.39}$, $\log_{10}P_\mathrm{cloudtop}$~[Pa]$>2.5$, $\log X_{\mathrm{H_2O}}=-1.36^{+0.11}_{-0.13}$, $\log X_{\mathrm{NH_3}}=-3.14^{+0.24}_{-0.26}$, $\log X_{\mathrm{CH_4}}=-1.64^{+0.16}_{-0.16}$, $\log X_{\mathrm{HCN}}=-6.12^{+1.98}_{-2.63}$, $\log X_{\mathrm{CO_2}}=-4.11^{+0.40}_{-0.45}$, $\log X_{\mathrm{CO}}=-6.69^{+2.14}_{-2.23}$, $\log X_{\mathrm{S}}=-3.56^{+0.21}_{-0.22}$, corresponding to 7--19$\times$ solar S/H.

This ultimately gives us [M/H] = $1.65^{+0.08}_{-0.10}$ (35--54$\times$solar) by summing up all the metals. These free retrievals suggested a sub-Rayleigh scattering haze with no evidence of a low-lying opaque cloud deck, consistent with the cloud/haze properties we found in semi-free retrieval in \S\ref{subsec:semi-free}.

\section{Grid retrieval results}
\label{sec:appendix_gridtrieval}


\begin{figure*}[ht!] 
    \centering 
    \includegraphics[width=\textwidth]{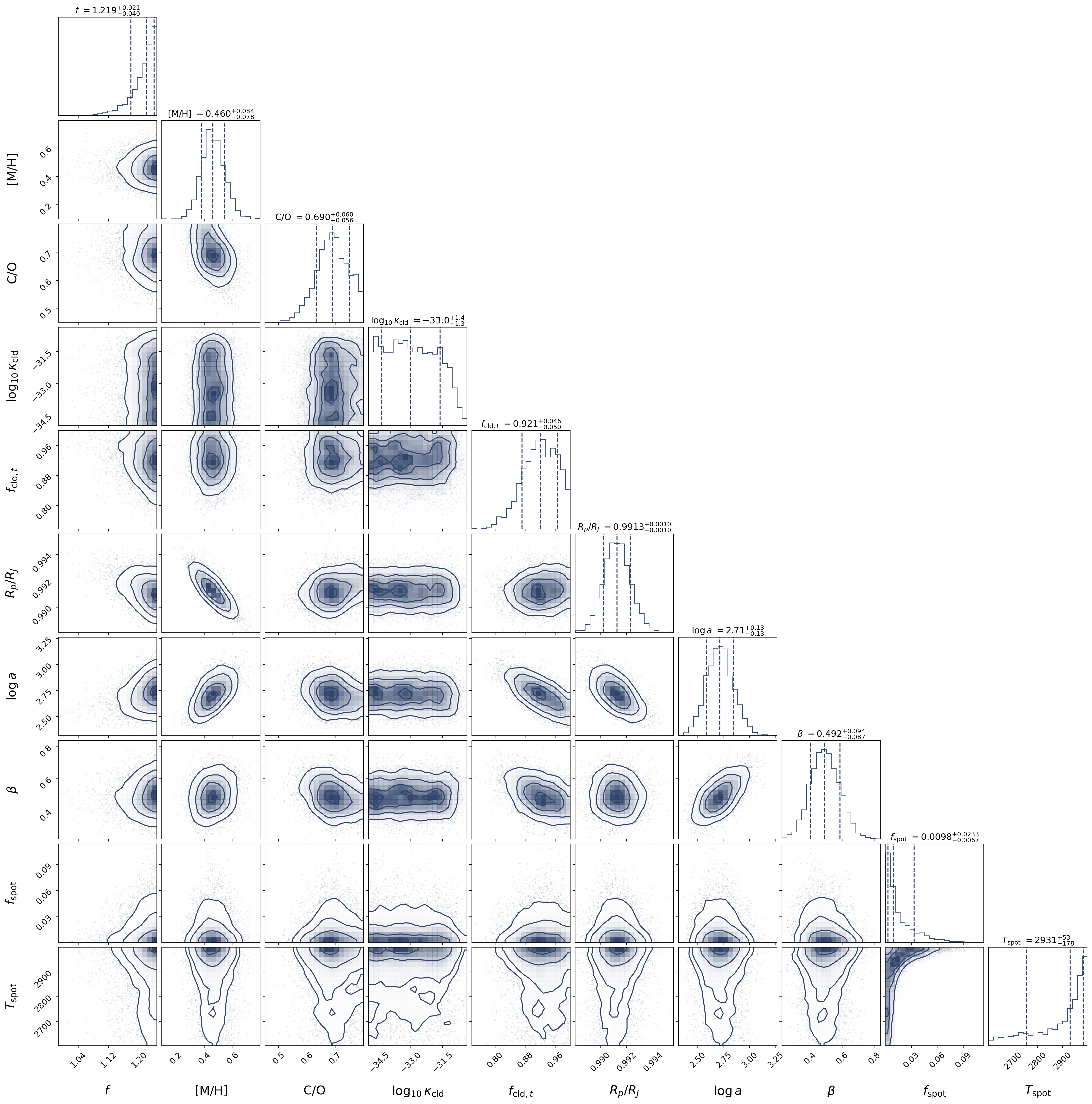}
    \caption{Corner plot from the ScCHIMERA + Photochem RCPE grid retrieval.  The red lines indicate grid points.}
    \label{fig:gridtrieval_corner}
\end{figure*}

The best fit from the grid retrieval is shown in Figure \ref{fig:spectra}, while the 2D posterior distributions (``corner plot'') is shown in Figure \ref{fig:gridtrieval_corner}.  $T_{\rm int}$, $K_{zz}$, and metal ratios other than C/O are fixed.

\bibliography{main}{}
\bibliographystyle{aasjournalv7}



\end{document}